\documentclass[aps,prb,twocolumn,superscriptaddress,notitlepage,showpacs]{revtex4-1}
\usepackage{graphicx}
\usepackage{times}
\usepackage{color}
\usepackage{subfigure}
\usepackage{float}
\usepackage[section]{placeins}
\usepackage{amsmath}
\usepackage{epstopdf}
\usepackage[colorlinks=true, urlcolor=blue, citecolor=blue, linkcolor=blue]{hyperref}

\usepackage[T1]{fontenc}
\usepackage[utf8]{inputenc}

\def  \bsig    {\mbox{\boldmath$\sigma$}}

\newcommand{\ii}{\mathrm{i}}

\newcommand{\doiit}[1]{
	{\href{http://dx.doi.org/#1}{http://dx.doi.org/#1}}
}

\begin{document}
\title{Topological insulator in a helicoidal magnetization field}
\date{\today}

\author{S. Stagraczy\'nski}
\affiliation{Department of Physics, Martin-Luther-Universit\"{a}t Halle-Wittenberg, 06120 Halle, Germany}

\author{L. Chotorlishvili}
\affiliation{Department of Physics, Martin-Luther-Universit\"{a}t Halle-Wittenberg, 06120 Halle, Germany}

\author{V. K. Dugaev}
\affiliation{Department of Physics and Medical Engineering, Rzesz\'ow University of Technology, 35-959 Rzesz\'ow, Poland}

\author{C.-L. Jia}
\affiliation{Key Laboratory for Magnetism and Magnetic Materials of MOE, Lanzhou University, Lanzhou 730000, China}

\author{A. Ernst}
\affiliation{Max-Planck-Institut für Mikrostrukturphysik, Weinberg 2, 06120 Halle, Germany}

\author{A. Komnik}
\affiliation{Institut für Theoretische Physik, Universität Heidelberg, Philosophenweg 12, D-69120 Heidelberg, Germany}

\author{J. Berakdar}
\affiliation{Department of Physics, Martin-Luther-Universit\"{a}t Halle-Wittenberg, 06120 Halle, Germany}

\begin{abstract} A key feature of topological insulators is the robustness of the electron energy spectrum. At a surface of a topological insulator, Dirac point is protected by the characteristic symmetry of the system. The breaking of the symmetry opens a gap in the energy spectrum. Therefore, topological insulators are very sensitive to magnetic fields, which can open a gap in the electronic spectrum. Concerning "internal" magnetic effects, for example the situation with doped magnetic impurities, is not trivial. A single magnetic impurity is not enough to open the band gap, while in the case of a ferromagnetic chain of deposited magnetic impurities the Dirac point is lifted. However, a much more interesting case is when localized magnetic impurities form a chiral spin order. Our first principle density functional theory calculations have shown that this is the case for Fe deposited on the surface of Bi$_2$Se$_3$ topological insulator. But not only magnetic impurities can form a chiral helicoidal spin texture. An alternative way is to use chiral multiferroics (prototype material is LiCu$_{2}$O$_2$) that induce a proximity effect. The theoretical approach we present here is valid for both cases. We observed that opposite to a ferromagnetically ordered case, a chiral spin order does not destroy the Dirac point. We also observed that the energy gap appears at the edges of the new Brillouin zone. Another interesting result concerns the spin dynamics. We derived an equation for the spin density dynamics with a spin current and relaxation terms. We have shown that the motion of the conductance electron generates a magnetic torque and exerts a certain force on the helicoidal texture.  \end{abstract}

\date{\today}

\maketitle

\section{Introduction}

The current interest to topological insulators (TI)\cite{Kane2010,Zhang2011} is twofold. Firstly, this is a new fascinating case study for theoretical approaches to understand "band-inverted" electronic structure \cite{Volkov85,Pankratov87} of semiconducting or insulating materials, based on the topology of corresponding wave functions and electron bands\cite{Fu2007}. Secondly, this is a new possibility to design materials and structures with a range of very unusual electronic and magnetic properties. One of such unusual properties is the rigidity of the electron energy spectrum at the TI surface in the vicinity of the Dirac point to any perturbation provided that they do not break the time-inversion symmetry, the property usually called the symmetry-protected Dirac point. Correspondingly, any potential-like perturbation (scattering) due to impurities does not destroy a topologically protected electronic structure, whereas ordered magnetic impurities can open a band gap \cite{Abanin2011,Efimkin2014,Wray2011,Sessi2014,Chotorlishvili2014}. The best studied case is that of the ferromagnetic ordering with a nonzero magnetization being perpendicular to the TI surface
\cite{KaneFu,Lutchyn,LutchynSarma}.
In some recent works the nontrivial role of the in-plane magnetic order has been discovered. In particular, it was found that the in-plane magnetic order
leads to an anomalous Josephson current \cite{BlackSchaffer,Dolcini}. Also, the nonconventional magnetization order can lead to some interesting physical phenomena. The purpose of our study is to consider the case when the magnetic ordering at the TI surface is noncollinear.

Usually, such magnetic ordering and spin frustrations occur due to the ferromagnetic nearest-neighbour exchange interaction competing with the next-nearest-neighbour antiferromagnetic coupling. Another source of noncollinear helicoidal magnetic order is the Dzyaloshinskii-Moriya (DM) interaction. The DM interaction appears due to the breaking of inversion symmetry at a surface. The helicoidal ordering of magnetic moments at the TI surface can be related to an indirect exchange interaction between the magnetic moments. As shown in Ref.~\onlinecite{Ye2010}, the exchange interaction mediated by electrons in  a TI contains the DM term due to the essential role of the spin-orbit interaction. This can lead to the formation of the helicoidal order of the moments. We checked this possibility performing Monte-Carlo simulations and first principle density functional calculations for a model of magnetic impurities (forming a chain at the surface of the Bi$_2$Se$_3$ topological insulator \cite{Zhang2009,Zhang2011}) and for a chiral multiferroic system. Our calculations show that due to the DM interaction, the spins of the impurities form an in-plane spiral magnetic texture. A particular feature of the in-plane spiral magnetic texture is the zero average magnetization component $\left< M_{z}\right> $.

The problem of our interest is not limited solely to magnetic impurities. Another possibility is to use a magnetic structure of a one-phase chiral multiferroic film placed on the top of a TI surface \cite{Li2011}. A remarkable advantage of the multiferroic materials is the strong magnetoelectric coupling, \cite{Schmid94,Tokura2010,Vaz2010,Zavaliche2007,Khomskii2006,Ramesh2007,Nan2008,Zhang2008,Scott2007}, which allows to control the magnetization via an external electric field. Moreover, in the case of multiferroics there exists a possibility to tune the period of the helicoid by an external electric field (since the period of the helicoid depends on the vector chirality, which can be modified via electric field).

The paper is organised as follows: In Sec.~II we describe a model of a 2D electron system on a surface of TI in the presence of a helicoidal magnetic field. This helicoidal magnetic field can be formed by a chiral multiferroic system (prototype material is  LiCu$_{2}$O$_2$);  an option, which is discussed in details in Sects.~III and IV. An alternative source for the  helicoidal magnetic field can be magnetic impurities forming a chain at the surface of the Bi$_2$Se$_3$ TI, which is discussed in Sec.~V. After studying the physics of localized spins we come back to the physics of conductance electrons. In Sections VI and VII, we study the energy spectrum of the system and spin dynamics of the conductance electrons. In Sec.~VIII we demonstrate that the mechanical torque generated by the motion of conductance electrons exerts a force on the helicoid formed by spins of the chain.

\section{Model with a continuous helicoidal magnetization}

We considered a model with TI electrons subject to a continuous helicoidal field generated by the magnetization ${\bf M}({\bf r})$ of some embedded magnetic subsystem (to be specified later). A continuous approximation for the helicoidal field is justified when the characteristic length of the variation of magnetization vector ${\bf M}$ is much larger then the distance $a$ between magnetic adatoms at the TI surface (or distance between lattice sites in the chiral multiferroic). We assume that in cartesian coordinates the magnetization field is described by
\begin{eqnarray}
	\label{eq:Mr}
	{\bf M}({\bf r})=\Big( M_0\cos ({\bf Q}_h \cdot {\bf r)},\; M_0\sin ({\bf Q}_h \cdot {\bf r)}, 0\Big)\,,
\end{eqnarray}
where ${\bf Q}_h$ is the wavevector of the helicoid. Then the condition of applicability of the continuous model is $Q_h a\ll 1$.

The Hamiltonian of 2D electrons at the surface of a TI coupled to this field reads
\begin{eqnarray}
	\label{eq:H}
	\hat{H}=\hat{H}_0+\hat{H}_{int}\,,
\end{eqnarray}
where
\begin{eqnarray}
	\label{eq:H0}
	\hat{H}_0=-\ii v\, (\sigma_x \partial _x+\sigma_y \partial _y)
\end{eqnarray}
describes the electron spectrum of the TI and
\begin{eqnarray}
	\label{eq:Hint}
	\hat{H}_{int}=g\, \bsig \cdot {\bf M}({\bf r}),
\end{eqnarray}
is the perturbation. Here $g$ is the coupling constant of the interaction of the TI electrons and the magnetization field ${\bf M}({\bf r})$.

Note that the complexity of this problem is related to the fact that for an arbitrary nonzero electron momentum
$\bf k\neq 0$ the two terms $\hat{H}_0$ and $\hat{H}_{int}$ do not commute. Therefore, the spin is not a good quantum number in this problem. The expectation value of the electron spin in TI with a helicoid magnetization  depends on the interaction constant $g$. A ferromagnetic ordering of the localized moments corresponds to a relatively simple case. In particular, an in-plane  ferromagnetic order ${\bf Q}_h=0$ leads to the spectrum $E=\pm \sqrt{v^{2}k^{2}+g^{2}M_{2}^{2}+2vgM_{0}k_{x}}$ and to the following shift of the Dirac point from $k_{x}=0,~k_{y}=0$ to  $k_{x}=-gM_{0}/v,~k_{y}=0$.

We choose the axis $x$ to be parallel to the vector ${\bf Q}_h$. Then  ${\bf Q}_h=(Q,\, 0)$ and $Q=2\pi /L$, where $L$ is the period of helicoid.

\begin{figure}[ht]
	\label{fig:model}
	\includegraphics[width=0.98\columnwidth]{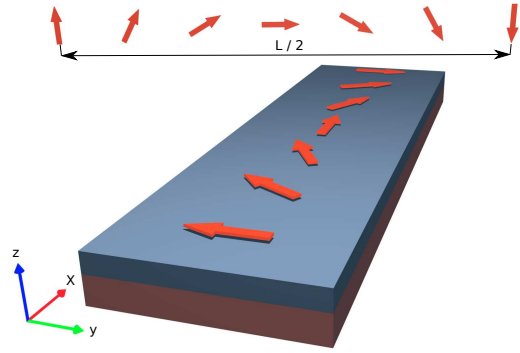}
	\caption{Schematic drawing of the helicoid orientation.}
\end{figure}

The eigenfunctions of the unperturbed Hamiltonian $\hat{H}_0$ are
\begin{eqnarray}
	\label{eq:psi_k}
	\psi_{{\bf k},\pm }({\bf r})= \frac{e^{\ii{\bf k\cdot r}}}{\sqrt{2\Omega }}\;
	\left(  \begin{array}{c} 1 \\  \pm \frac{k}{k_{-}} \end{array} \right)\,,
\end{eqnarray}
where $k_{\pm} = k_x \pm \ii k_y$. They correspond to the energy branches with linear dispersion, $\varepsilon _\pm ({\bf k})=\pm vk$.

The periodic along axis $x$ perturbation (\ref{eq:Hint}) affects the energy spectrum and eigenfunctions of the TI. It is known that a homogeneous magnetization field can open a gap in the energy spectrum of Dirac electrons. We can show that this is not the case of inhomogeneous field (\ref{eq:Mr}) with a zero average in space magnetization, $\left< {\bf M(r)}\right> =0$.

The formation of a helical spin order in the system of localized spins (either addatoms or chiral multiferroic) deserves a particular attention and is presented in the next section.

\section{Helicoidal order in the system of localized spins}

In this section we rigorously deduce the helicoidal magnetization field ${\bf M}({\bf r})$ from the spin chain model of localized moments. One way to generate a periodic magnetization of the structure given in Eq.~(\ref{eq:Mr}) is with the help of a frustrated spin chain with a non-zero next nearest neighbour (NNN) coupling in addition to the conventional nearest neighbour exchange. The minimal Hamiltonian for the localized spins then reads
\begin{eqnarray}
\label{eq:FrustratedHeisenberg} \hat{H}_{chain} = J_1 \sum\limits_{i=1}^{N} {\bf S}_i \cdot {\bf S}_{i+1} + J_2 \sum\limits_{i=1}^{N} {\bf S}_i \cdot  {\bf S}_{i+2} \nonumber\\
- \boldsymbol{\mathcal{E}} \cdot {\bf P} \, ,
\end{eqnarray}
where the coupling parameters $J_{1,2}$ usually carry opposite signs. Typical values, e.g., for LiCu$_{2}$O$_2$ are: $J_{1}=-11\pm 3$ meV and $J_{1}=+7\pm 1$ meV, see Refs.~\onlinecite{Park2007,Schrettle2008}. Opposite signs of the exchange constants usually lead to  spin frustrations and to the formation of a chiral spin order. We also allow the chain to be ferroelectrically active adding a term proportional to the external electric field $\boldsymbol{\mathcal{E}}$, which couples to the ferroelectric polarization vector
\begin{eqnarray}
	\label{eq:P}
	{\bf P} = g_{ME} \sum\limits_{i=1}^{N} \left[ {\bf e}_x \times \left( {\bf S}_i \times {\bf S}_{i+1} \right)\right]
\end{eqnarray}
with the amplitude of magnetoelectric coupling $g_{ME}$.

% At zero electric field ${\mathcal{E}}=0$
 In the absence of the electric field $\boldsymbol{\mathcal{E}}$, the system is of the celebrated Majumdar-Ghosh type and possesses very interesting properties.\cite{MajumdarGhosh1967,Chubukov91,Bursill1995} They have been studied in detail in the context of frustrated two-leg zig-zag spin ladders. Among the most spectacular findings is for example the possibility of a spiral incommensurate order.\cite{Nersesyan1998} Applying a finite magnetic field or breaking the $SU(2)$ symmetry of the couplings (for instance introducing a coupling anisotropy) leads to a non-zero expectation value of the
 vector chirality
 \begin{eqnarray}
  \boldsymbol{\kappa}= ( {\bf S}_{i} \times {\bf S}_{i+1} ) \, ,
 \end{eqnarray}
 (see e.~g. [\onlinecite{Kecke2008},\onlinecite{Sirker2010}]), and thus to a spin ordering, the $xy$-projection of which is given in Eq.~\eqref{eq:Mr}. Obviously, the same effect takes place when $\boldsymbol{\kappa}$ appears as a perturbation of a Hamiltonian explicitly as it is the case for  $g_{ME} \neq 0$. The ground state is easiest to understand in the case of switched off NNN coupling, when $J_2=0$. For simplicity, we assume the electric field $\boldsymbol{\mathcal{E}}$ to be oriented along the $y$-axis, then the magnetoelectric coupling term is
 \begin{eqnarray}
   \boldsymbol{\mathcal{E}} \cdot {\bf P} = {\mathcal{E}} g_{ME} (S^x_{i} \, S^y_{i+1} - S^y_{i} \, S^x_{i+1}) \, .
 \end{eqnarray}
 Formally it has precisely the same form as the Dzyaloshinskii-Moriya (DM) interaction along the $z$-direction. It is known that in this case the $J_2=0$ isotropic chain can be mapped onto an anisotropic Heisenberg chain with the Hamiltonian
 \begin{eqnarray}
  \hat{H} = -J' \sum\limits_{i=1}^{N} {\bf S'}_i \cdot {\bf S'}_{i+1} + \Delta S'^z_i \, S'^z_{i+1} \, ,
 \end{eqnarray}
 where ${S}_j^\pm = e^{\pm i j \theta} {S'}_j^\pm$, $S_j^z = {S'}_j^z$, $\theta = - \arctan ({\mathcal{E}}g_{ME}/J)$, $J' = J/ \cos \theta $, and $\Delta = \cos \theta $.\cite{Bocquet2001} Thus the ground state of our model can be mapped out from the ground state of the anisotropic Heisenberg chain by a simple spatial `twist' with an angle $\theta$ between the adjacent sites.

The key point is that angle between adjacent spins continuously depends on the electric field $\theta = - \arctan ({\mathcal{E}}g_{ME}/J)$. The period of the helicoid reads $\theta (n-m)=2\pi$, where $n$,$m$ are the site numbers. Using notations of the (\ref{eq:Mr}) we deduce:  ${\bf Q}_h\cdot {\bf r}_l=2\pi= (m-n)\arctan ({\mathcal{E}}g_{ME}/J)$. Here $|{\bf r}_l|=L$ is the length of the helicoid.

Switching on finite $J_2$ potentially changes the induced ${\bf M}$ slightly as in this case the $\theta$ periodicity starts to compete with the incommensurability effects due to frustrations. However localized spins still form a helicoidal structure. A possible way to understand this might be a generalization of the approach presented in [\onlinecite{Nersesyan1998}]. To support of the theoretical estimations, we performed Monte Carlo calculations, which  are presented in the next section.

\subsection{Monte Carlo simulation}

To support to the analytical estimates we performed a Monte Carlo study of the Hamiltonian (\ref{eq:FrustratedHeisenberg}) for the magnetic moments.
%Such effective model Hamiltonian is relevant for 1D spin frustrated MF oxides, e.g. LiCu$_{2}$O$_2$. Typical values of the relevant parameters are  $-J_{1}\approx J_{2}$; e.g. for LiCu$_{2}$O$_2$ one finds \cite{Park2007,Schrettle2008} $J_{1}\approx -11 \pm 3 $ meV and $J_{2}\approx 7 \pm 1 $ meV.
A competition between the nearest-neighbour ferromagnetic coupling with the next-nearest-neighbour antiferromagnetic coupling leads to the chiral spin structure with the pitch angle $\arccos(-J_{1}/4J_{2})$ for $J_{2} > |J_{1}|/4$ at $T=0$. In Monte Carlo calculations we adopt dimensionless units $J_2 \rightarrow J_2 / |J_1|$, $k_{B}T \rightarrow k_{B}T/|J_1|$, and set the parameters $-J_1 = J_2 = 1$.

The spiral plane of frustrated spins is found to be sensitive to the applied electric field due to the coupling term
\begin{eqnarray}
	\boldsymbol{\mathcal{E}} \cdot {\bf P} = {\mathcal{E}} g_{ME} \sum\limits_{i=1}^{N} \left[ {\bf e}_x \times \left( {\bf S}_i \times {\bf S}_{i+1} \right)\right].
\end{eqnarray}
The electric field along Y-axis $\boldsymbol{\mathcal{E}} = (0,\mathcal{E}_y,0)$ stabilizes the ferroelectric polarization $\boldsymbol{P}$ in parallel.  Fig.~2 shows that the maximum of the spin structure $\langle S_{k}\cdot S_{-k} \rangle$ tends to diverge at T=0, indicating the formation of  a helical long-range magnetic order  given in Eq.~(\ref{eq:Mr}). The expectation values of the spins $\left\langle {\bf S}_i \right\rangle$ can be evaluated via Monte Carlo method as well.
%:
%\begin{eqnarray}
%	{\bf P} = g_{ME} \sum\limits_{i=1}^{N} \left[ {\bf e}_x \times \left( {\bf S}_i \times {\bf S}_{i+1} \right)\right]
%\end{eqnarray}
%If $z$ component of the vector chirality is non-zero
%\begin{eqnarray}
%	{\kappa_{z}}=\left\langle \sum\limits_{i=1}^{N} \left[ {\bf e}_x \times \left( {\bf S}_i \times {\bf S}_{i+1} \right) \right]_z \right\rangle \neq 0
%\end{eqnarray}
%due to the coupling term
%\begin{eqnarray}
%	\boldsymbol{\mathcal{E}} \cdot {\bf P} = {\mathcal{E}} g_{ME} \sum%\limits_{i=1}^{N} \left[ {\bf e}_x \times \left( {\bf S}_i \times {\bf S}_{i+1} \right)\right]
%\end{eqnarray}
%magnetic order becomes sensitive to the electric field. The expectation values of the spins $\left\langle {\bf S}_i \right\rangle$ can be evaluated via Monte Carlo method.
\begin{figure}[ht]
	\centering
	\includegraphics[width=0.98\columnwidth]{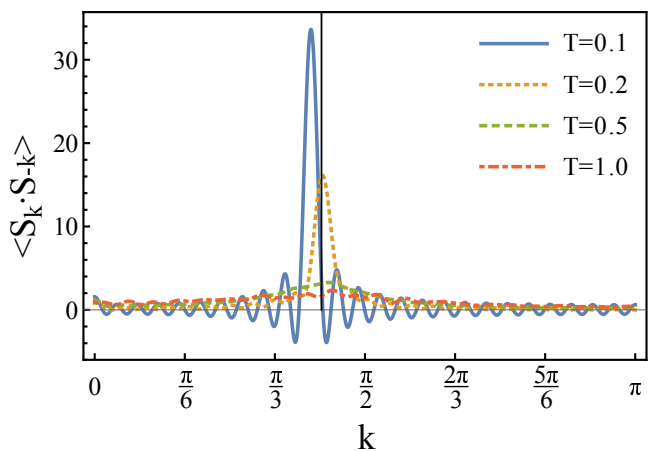}
	\caption{The static spin structure factor $\left< S_k \cdot S_{-k}\right> $ for different temperatures. Fourier-transformed correlation function is
	$\left<S_k \cdot S_{-k}\right> = 1 + \sum\limits_{n=1}^{M} \left< S_0 \cdot S_n \right> \cos k n$ and number of spins $M=100$.}
	\label{fig:SxSx}
\end{figure}
%\begin{figure}[hb]
%	\includegraphics[width=0.98\columnwidth]{f_Si2.pdf}
%	\caption{Taken from MC simulation.}
%\end{figure}

%\section{$K$-$p$ approach: low-energy solution}
%Calculations confirmed formation of the helicoidal order in the system of localized spins.

\subsection{Noncollinear helicoidal magnetic system: $\mathrm{Fe}$ in $\mathrm{Bi}_2 \mathrm{Se}_3$}

%A possible candidate as a system with a noncollinear magnetic order can be suggested a Bi$_2$Se$_3$ (0001) surface doped with Fe. A such material can be fabricated by deposing a certain amount of Fe on the  Bi$_2$Se$_3$ (0001) surface\cite{Polyakov2015}.
A possible candidate as a system with a noncollinear magnetic order can be a Bi$_2$Se$_3$ (0001) surface doped with Fe. Such material can be fabricated by deposing a certain amount of Fe on the  Bi$_2$Se$_3$ (0001) surface\cite{Polyakov2015}. Therewith, Fe atoms replace Bi in the first quintuple layer. The impurity concentration can be controlled during the experiment. Using the experimentally found geometry, we performed a first-principles study on electronic and magnetic structures of (Bi$_{1-x}$Fe$_x$)$_2$Se$_3$/Bi$_2$Se$_3$ (0001) using a self consistent full relativistic Green function method~\cite{Geilhufe2015} designed to treat semi-infinite systems such as surfaces and interfaces~\cite{Luders2001}. The calculations were carried out within the density functional theory in a generalized gradient approximation~\cite{Perdew1996}. To simulate Fe impurities in the Bi$_2$Se$_3$ layer a coherent potential approximation~\cite{Soven1967} was used as it is implemented within the multiple scattering theory~\cite{Gyorffy1972}. Noncollinear magnetic configurations were calculated using a noncollinear magnetism approach within the density functional theory~\cite{Sandratskii1998}.

\begin{figure}[ht]
	\centering
	\includegraphics[width=0.98\columnwidth]{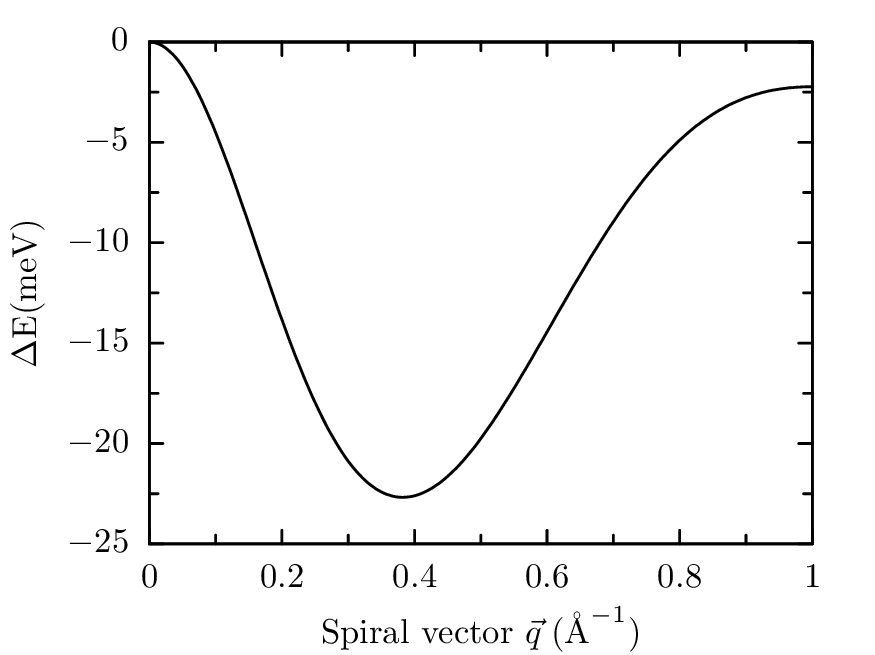}
	\caption{ Total energy difference as a function of the spiral vector ${\bf q}_\parallel$ in $\bar{\Gamma}$-$\bar{K}$ direction.}
	\label{fig:Delta_E}
\end{figure}

\begin{figure}[ht]
	\centering
		\includegraphics[width=0.98\columnwidth]{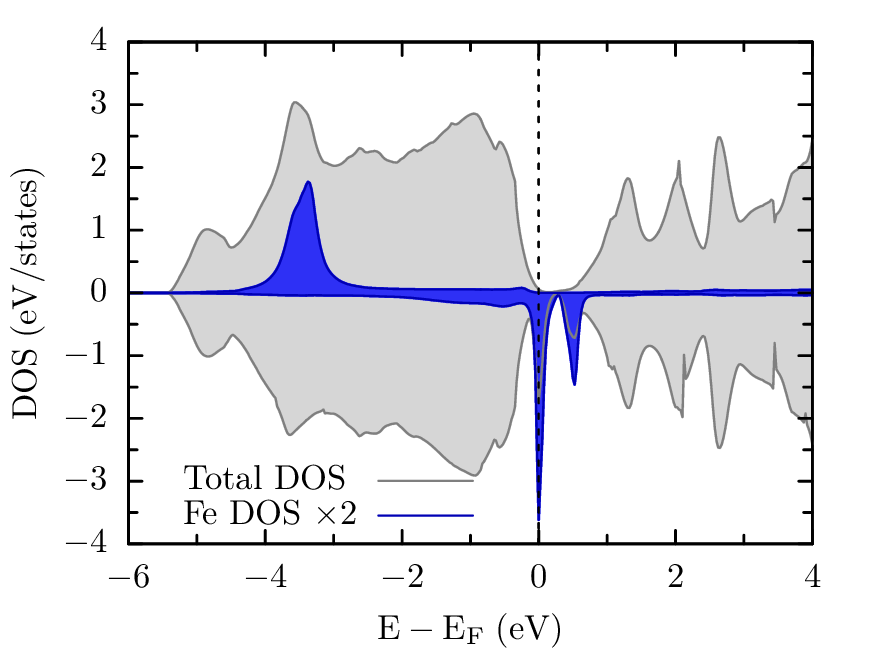}
	\caption{Total spin-resolved density of states of the surface quintuple layer and that of Fe impurities in (Bi$_{0.95}$Fe$_{0.05}$)$_2$Se$_3$/Bi$_2$Se$_3$~(0001). The DOS of Fe is scaled by the factor 2. The negative DOS here means the DOS of the minority spin channel.}
	\label{fig:dos}
\end{figure}

Figure~\ref{fig:Delta_E} shows the calculated total energy difference of (Bi$_{0.95}$Fe$_{0.05}$)$_2$Se$_3$/Bi$_2$Se$_3$~(0001) as a function of the spiral vector ${\bf q}_\parallel$ in  $\bar{\Gamma}$-$\bar{K}$ direction. The total energy minimum was found  at ${\bf q}_\parallel$=\{0.4,0.0\} $\AA^{-1}$. A large exchange splitting of 3.5 eV and very narrow $3d$ Fe states are responsible for a such magnetic behaviour (see the corresponding density of states on Fig.~\ref{fig:dos}). Fe atoms possess a magnetic moment of 3.4 $\mu_{B}$ and interact with each other with the exchange energies of 1.5 meV and -1.3 meV for the first and the second neighbour shells, respectively. A such competing exchange interaction between the next-nearest-neighbour results in a noncollinear helical magnetic order.

\section{The $k \cdot p$ approach: low-energy solution}
	
Let us consider now the electron energy spectrum of the TI with magnetic helicoidal order (1) using Hamiltonian (2).
In a close vicinity to the Dirac point ${\bf k}=0$ one can use the $k \cdot p$ perturbation method. The Schr\"odinger equations for the spinor components $\varphi (x),\chi (x)$ of the wavefunctions $\psi ^{T}_{\bf k}({\bf r})=e^{\ii{\bf k\cdot r}}\left( \varphi ,\chi \right)$ of the perturbed Hamiltonian $\hat{H}$ are
\begin{eqnarray}
	\label{eqs:Schro_a}&&-\varepsilon \varphi +\left( -\ii v\partial _x+vk_-+gM_0e^{-\ii Qx}\right) \chi =0,\\
	\label{eqs:Schro_b}&&\left( -\ii v\partial _x+vk_++gM_0e^{\ii Qx}\right) \varphi -\varepsilon \chi =0\,.
\end{eqnarray}
Let us assume that for $k_x=k_y=0$ there is a solution of Eqs.~(\ref{eqs:Schro_a}),(\ref{eqs:Schro_b}) with the energy $\varepsilon =0$. In this case Eqs.~(\ref{eqs:Schro_a}) and (\ref{eqs:Schro_b}) acquire the following form
\begin{eqnarray}
	\label{eqs:Schro_k0_a} &&-\ii v\chi '+gM_0e^{-\ii Qx}\chi =0\,,\\
	\label{eqs:Schro_k0_b} &&-\ii v\varphi '+gM_0e^{\ii Qx}\varphi =0\,,
\end{eqnarray}
and we obtain the solution for $\varepsilon =0$
\begin{eqnarray}
	\label{eqs:sol_phi}	\varphi (x)&=& C_1\exp \left( -\frac{gM_0e^{\ii Qx}}{Qv}\right),\\
	\label{eqs:sol_chi}	\chi (x)&=& C_2\exp \left( \frac{gM_0e^{-\ii Qx}}{Qv}\right)\,.
\end{eqnarray}
Then following the $k \cdot p$ method we take two basis functions
\begin{eqnarray}
	\label{eqs:basis_up} \psi _{1{\bf k}}({\bf r})&=& C_1e^{\ii{\bf k\cdot r}} \left( \begin{array}{c} 1 \\ 0 \end{array} \right)\exp \left( -\frac{gM_0e^{\ii Qx}}{Qv}\right)\,,\\
	\label{eqs:basis_down} \psi _{2{\bf k}}({\bf r})&=& C_2e^{\ii{\bf k\cdot r}} \left( \begin{array}{c} 0 \\ 1\end{array} \right) \exp \left( \frac{gM_0e^{-\ii Qx}}{Qv}\right)\,,
\end{eqnarray}
where $C_1$ and $C_2$ are constants determined by normalization.
%of the basis functions (\ref{eqs:basis_up}) and (\ref{eqs:basis_down}).

The matrix of the Hamiltonian $\hat{H}$ in the basis of functions (\ref{eqs:basis_up}) and (\ref{eqs:basis_down}) is
\begin{eqnarray}
	\label{eq:H_kp}
	\hat{H}_{k\cdot p}=\left( \begin{array}{cc} 0 & \tilde{v}k_- \\ \tilde{v}^*k_+ & 0 \end{array} \right)\,,
\end{eqnarray}
where
\begin{eqnarray}
	\label{eq:v}
	\tilde{v}=v\, C_1C_2\mathcal{L}_y\int _0^{\mathcal{L}_x}dx\, \exp \left( \frac{2gM_0e^{\ii Qx}}{Qv}\right)\,.
\end{eqnarray}
$\mathcal{L}_x$ and $\mathcal{L}_y$ are the sizes of the sample in $x$ and $y$ directions, and we use a normalization of the wavefunctions $\psi _{1,2\, {\bf k}}({\bf r})$ in a 2D box with dimensions $\mathcal{L}_x\times \mathcal{L}_y$. Obviously, it follows from (\ref{eq:v}) that in the present $k \cdot p$ approximation the energy spectrum near the Dirac point is defined by the renormalized electron velocity $\tilde{\varepsilon }_\pm({\bf k})=\pm |\tilde{v}|k$.
%and the constant $\tilde{v}$ corresponds to the renormalized velocity of electrons near ${\bf k}=0$.

As we see, there is no energy gap in the Dirac point. This is related to the property of Hamiltonian (\ref{eq:H}) with respect to the unitary transformation $\sigma _z\hat{H}\sigma _z=-\hat{H}$, which means that if a function $\psi ({\bf r})$ is a solution of Schr\"odinger equation for the energy $\varepsilon $ then $\sigma _z\psi ({\bf r})$ is the solution for the energy $-\varepsilon$. Correspondingly, the existence of a solution with $\varepsilon =0$ implies the existence of another with the property  $\psi ({\bf r})=\sigma _z\psi ({\bf r})$. This is obviously fulfilled for the functions (\ref{eqs:basis_up}) and (\ref{eqs:basis_down}). The solution we found describes the wave functions and the energy spectrum of the states near the point ${\bf k}=0$.

\section{Perturbation theory}

Now we calculate the energy spectrum in the periodic perturbation $\hat{H}_{int}({\bf r})$ not restricting ourselves by a small neighbourhood of the Dirac point. For this purpose we use the perturbation theory. Due to the periodicity of $\hat{H}_{int}({\bf r})$, the only nonzero matrix elements of this perturbation couple an arbitrary state with wavevector ${\bf k}$ to the states with ${\bf k\pm Q}$.
Then, assuming  a weak perturbation $g/vQ \ll 1$ we can use the basis of only six functions. They are eigenfunctions (\ref{eq:psi_k}) of the Hamiltonian $\hat{H}_0$ corresponding to the wavevectors ${\bf k}$ and $\bf k\pm Q$
%\begin{eqnarray}
%	\label{eqs:phi_k}
%	\Phi_{1{\bf k}}&=& \psi_{{\bf k},+ },\; \Phi_{2{\bf k}}= \psi_{{\bf k+Q},+ },\; \Phi_{3{\bf k}}= \psi_{{\bf k-Q},+ },\nonumber \\
%	\Phi_{4{\bf k}}&=& \psi_{{\bf k},- },\; \Phi_{5{\bf k}}= \psi_{{\bf k+Q},- },\; \Phi_{6{\bf k}}= \psi_{{\bf k-Q},- }.
%\end{eqnarray}
\begin{align}
	\label{eqs:phi_k}
	\Phi_{1{\bf k}}&= \psi_{{\bf k},+ },\; \Phi_{2{\bf k}}= \psi_{{\bf k+Q},+ },\; \Phi_{3{\bf k}}= \psi_{{\bf k-Q},+ },\nonumber \\
	\Phi_{4{\bf k}}&= \psi_{{\bf k},- },\; \Phi_{5{\bf k}}= \psi_{{\bf k+Q},- },\; \Phi_{6{\bf k}}= \psi_{{\bf k-Q},- }\,.
\end{align}

\begin{widetext}
The Hamiltonian $\hat{H}$ in this basis is a matrix
	\begin{gather}
		  \hat{H}_{per} =
		  \left(
		   \begin{array}{lcclcc}
		   \ \ \ v|{\bf k}| & A^*   & B     & \ \ \ 0 & -A^*   & B		\\
		   \ \ \ A	 & v|{\bf k+Q}| & 0     & \ \ \ A & 0      & 0		\\
		   \ \ \ B^* & 0     & v|{\bf k-Q}| & -B^*    & 0      & 0		\\
		   \ \ \ 0	 & A^*   & -B    & -v|{\bf k}|    & -A^*   & -B	\\
		     	-A   & 0     & 0     & -A      & -v|{\bf k+Q}| & 0		\\
		   \ \ \ B^* & 0     & 0     & -B^*    & 0      & -v|{\bf k-Q}|\\
		   \end{array}
		  \right)\,, \nonumber \\
	 	 \label{mat:Hper}
	\end{gather}
\end{widetext}

where we denote
\begin{eqnarray}
	\label{eq:mat_elem}
	 A = \frac{gM_0\, |{\bf k+Q}|}{k_+ + Q}\, ,\hskip0.5cm  B=\frac{gM_0\, |{\bf k}|}{k_+}\, .
\end{eqnarray}

The eigenfunctions of the full Hamiltonian $\hat{H}$, Eq.~(\ref{eq:H}), can be written down as a superposition of the basis functions
\begin{eqnarray}
	\label{eq:psi_nkr}
	\psi_{n{\bf k}}({\bf r})=\sum _{m=1}^6 c_{nm}\Phi _{m{\bf k}}({\bf r})\,,
\end{eqnarray}
where ${\bf c}_n=(c_{n1},...,c_{n6})$ is an eigenvector of the matrix (\ref{mat:Hper}) corresponding to the energy $\varepsilon _{n{\bf k}}$ in the $n$-th energy band.

\subsection{Electron energy bands}

\begin{figure}[!ht]
		\includegraphics[width=0.98\columnwidth]{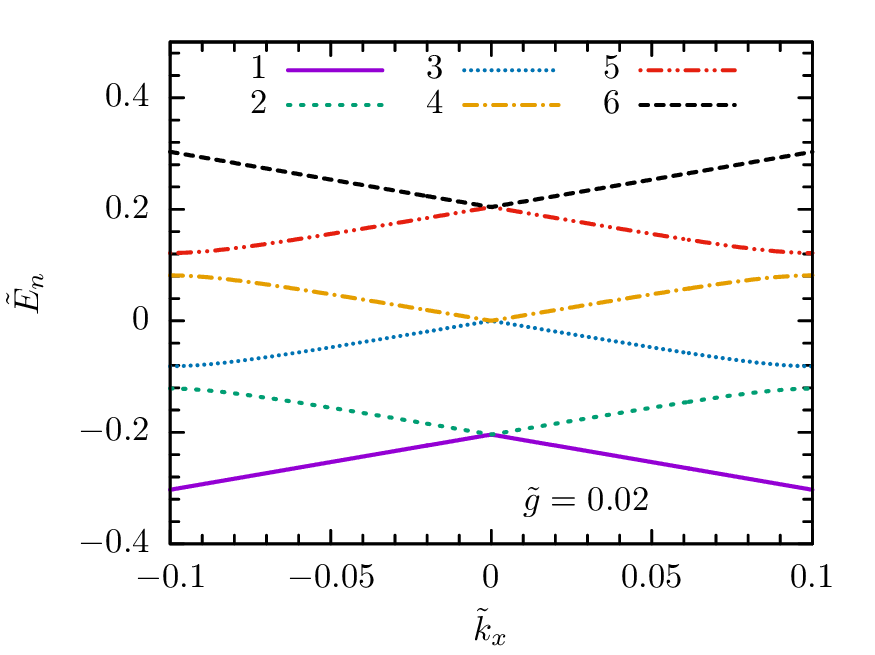}
		\caption{The energy spectrum $\tilde{E}_n(k_x)$ for fixed values of $\tilde{Q}=0.2$, $\tilde{g}=0.02$,
		$\tilde{k}_y=0.0$. The Brillouin zone edges are at $k_x=\pm Q/2$, which corresponds to $\tilde{k}_x=\pm 0.1$. }
		\label{fig:Ekx}		
	\end{figure}

\begin{figure}[!ht]
		\includegraphics[width=0.98\columnwidth]{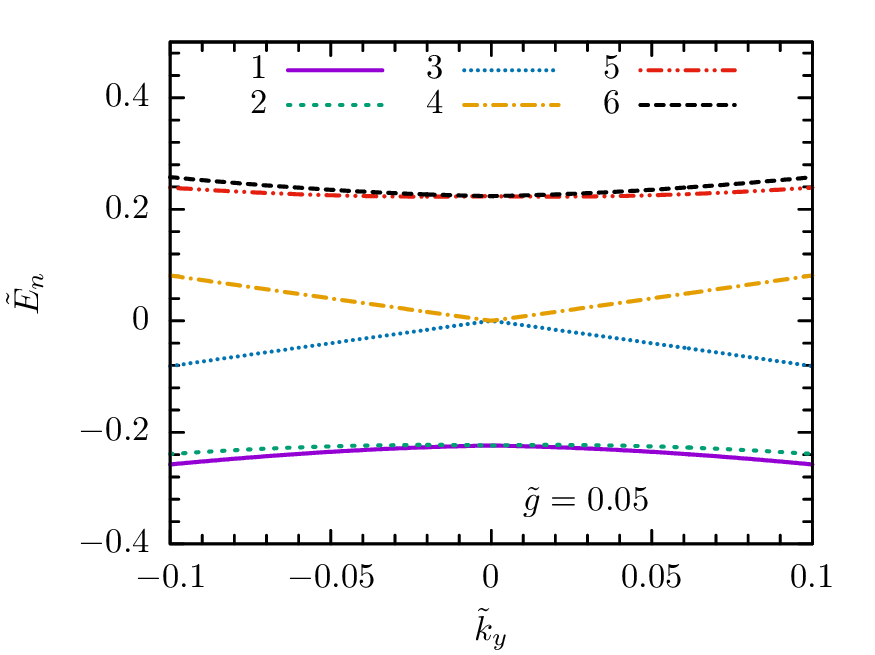}
		\caption{The energy spectrum $\tilde{E}_n(k_y)$ for fixed values of $\tilde{Q}=0.2,\ \tilde{g}=0.05, \tilde{k}_x=0.0$.}
		\label{fig:Eky}		
	\end{figure}

The electron energy bands, calculated numerically by using the Schr\"odinger equation with the matrix (\ref{mat:Hper}), are presented in Figs.~\ref{fig:Ekx} and \ref{fig:Eky} as a function of $k_x$ and $k_y$, respectively. We use dimensionless energies and coupling strengths and denote  $\tilde{k}=k/k_0$, $\tilde{E}_{n{\bf k}}=E_{n{\bf k}}/vk_0$, $\tilde{g}=gM_0/vk_0$ and $\tilde{Q}=Q/k_0$, where $k_0$ can be chosen arbitrarily. One can take $k_0=10^5$~cm$^{-1}$, and then we get $gM_{0}=\tilde{g}vk_{0}=0.02$~meV.

As we see in Fig.~\ref{fig:Ekx}, the energy gap is zero at ${\bf k}=0$ but the gap appears at the edges of the Brillouin zone $k_x=\pm Q/2$. However, since the periodicity of the perturbation is only along the axis $x$, there is no gap in $k_y$-direction, which means that electrons can continuously fill the states with larger values of energy $\varepsilon $ by growing $k_y$. But there is the gap at the edges of the Brillouin zone at $k_x=\pm Q/2$ for any value of $k_y$ including the point $k_y=0$. This is because the perturbed Hamiltonian (2), which is not commuting with $\sigma _y$ for any $k_y$, so that the equations for spinor components of the wavefunction are always coupled. This leads to an effective periodic perturbation in equations for the spinor components similar to the case of usual electron gas with a periodic potential along axis $x$.

\subsection{Spin polarization}

Using the wavefunctions (\ref{eq:psi_nkr}) of the perturbed Hamiltonian one can find the spin polarization of the electrons in the state ($n,{\bf k}$). Without perturbation related to the magnetization (\ref{eq:Hint}), the eigenstates of TI Hamiltonian $\hat{H}_0$ are fully spin-polarized along the vector ${\bf k}$ or $-{\bf k}$ (i.e. they are the states with a certain helicity). Due to the perturbation (\ref{eq:Hint}), the spin is no longer a good quantum number but the average value of the spin polarization ${\bf S}_{n{\bf k}}=\left< n{\bf k}|\bsig |n{\bf k}\right> $ is not zero and is depending on $n$ and ${\bf k}$.
The matrices of the spin operator components in the basis of the functions (\ref{eqs:phi_k}) are listed in appendix \ref{app:S}.%have the following form	

	\begin{figure}[!ht]
		\includegraphics[width=0.98\columnwidth]{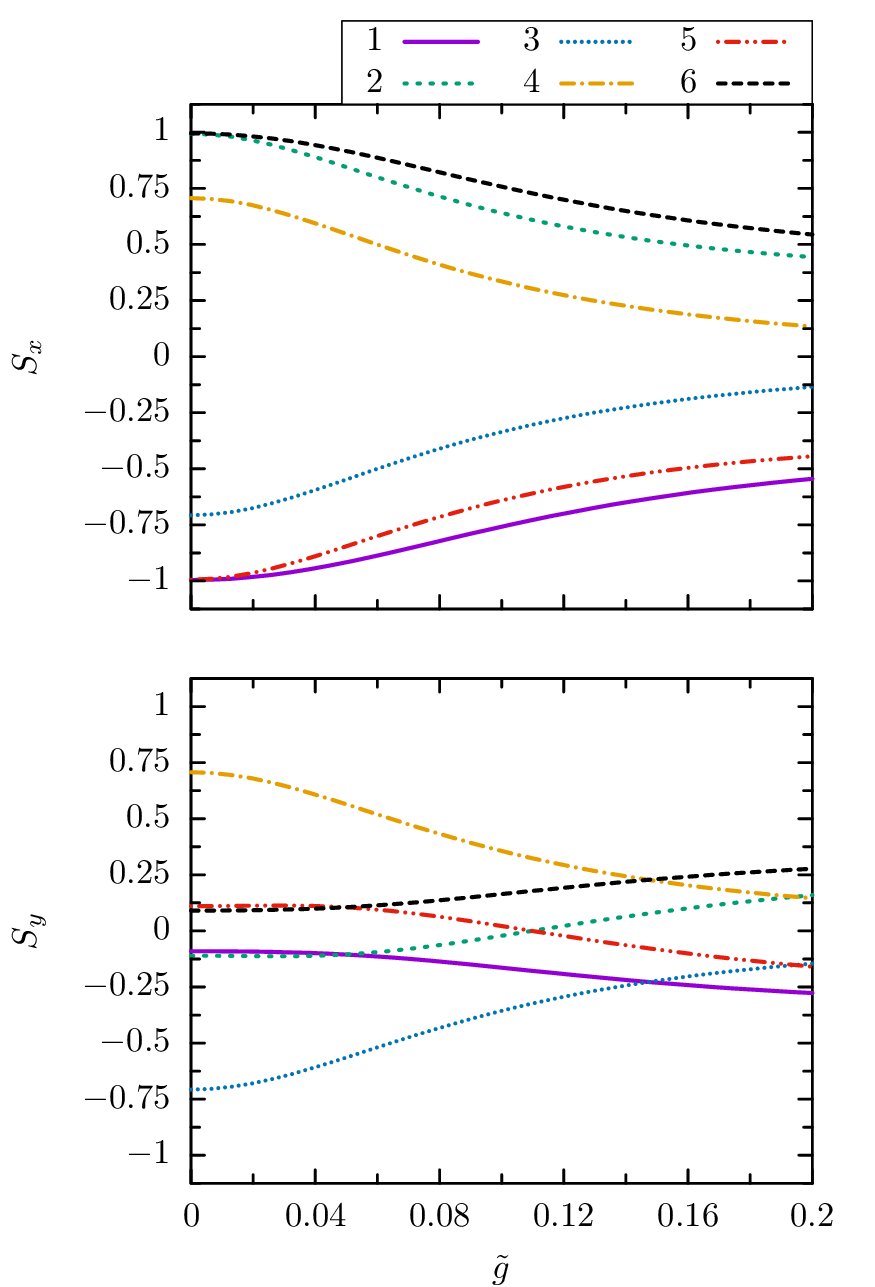}
		\caption{Mean values of spin polarization along axes $x$ and $y$ as a function of coupling constant $\tilde{g}$ for $\tilde{k}_x = \tilde{k}_y = 0.02$ and $\tilde{Q} = 0.2$.}
		\label{fig:SxSy(g)}		
	\end{figure}

\begin{figure}[!ht]
		\includegraphics[width=0.98\columnwidth]{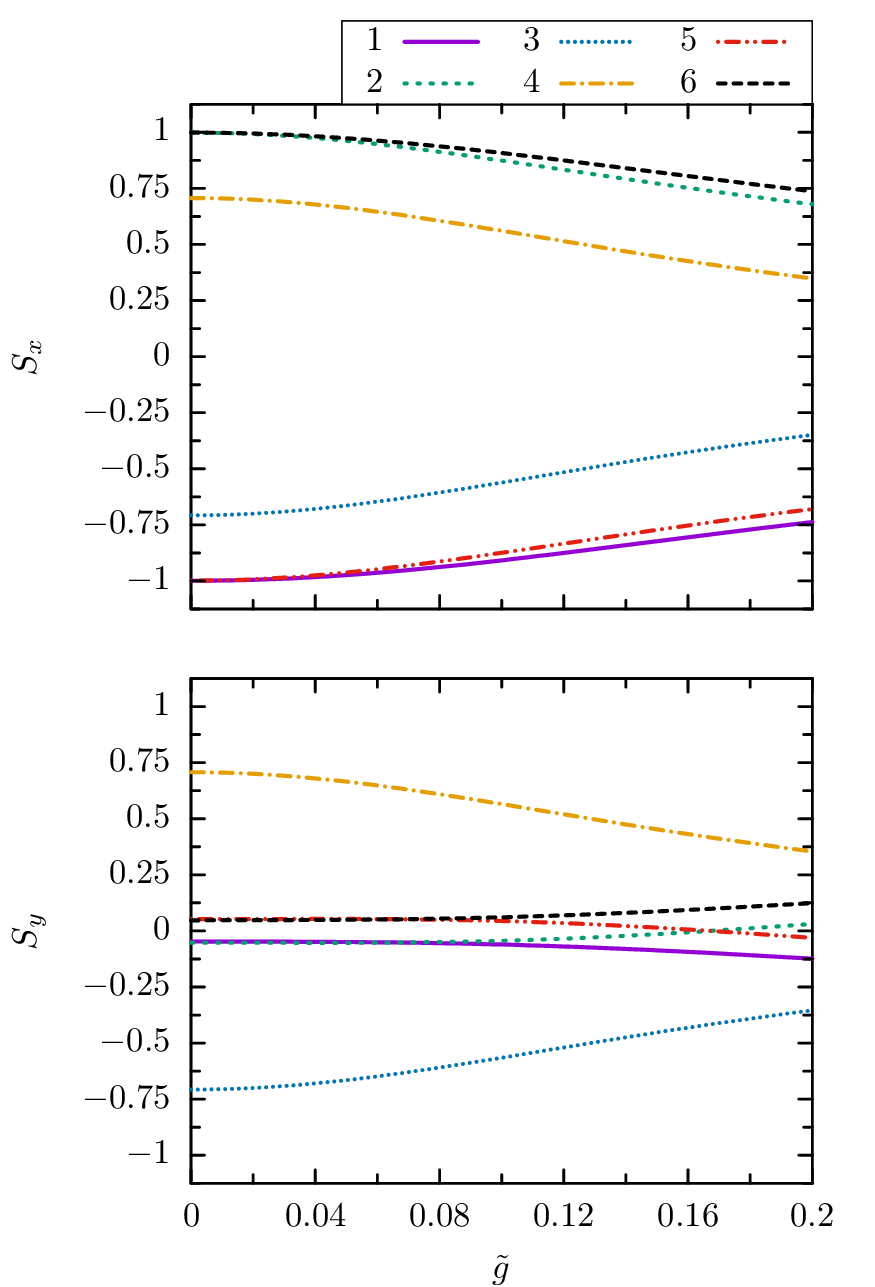}
		\caption{Mean values of spin polarization along axes $x$ and $y$ as a function of coupling constant $\tilde{g}$ for $\tilde{k}_x = \tilde{k}_y = 0.02$ and $\tilde{Q} = 0.4$.}
		\label{fig:SxSyQ4(g)}		
	\end{figure}
	
Then, the average value of spins can be calculated numerically by using Eqs.~(\ref{mat:Sxk}),(\ref{mat:Syk}) with eigenvectors of the matrix (\ref{mat:Hper}). This value characterizes a spin polarization of an eigenstate of the Hamiltonian $\hat{H}$.  The average values $S_x$ and $S_y$ components are presented in Figs.~\ref{fig:SxSy(g)} and \ref{fig:SxSyQ4(g)} as a function of coupling constant for $\tilde{Q}=0.2$ and $\tilde{Q}=0.4$, respectively. As we see, the mean spin polarization of electrons is nonzero but decreases with a growing coupling strength to the magnetization.  Comparing the results in these two figures, we see that the larger is the transferred momentum $\tilde{Q}$, the smaller are the expectation values of the in-plain spin components. The result is comprehensible as a  stronger scattering leads to a stronger deviation from the ground state spin configuration (the spin collinear to the wave vector).
Using Eq.~(\ref{mat:Szk}) we also found that the mean value of the spin polarization along axis $z$ is zero.

\subsection{Spin dynamics}

The dynamics of the electron spin density $S_\mu =\psi ^\dag \sigma _\mu \psi $ can be described by using the standard equation for spin density variation\cite{spindensity}
\begin{eqnarray}
	\label{eq:std_spin_density}
	\frac{\partial }{\partial t}\left( \psi ^\dag \sigma _\mu \psi \right)
	=\frac{\ii}{\hbar}\, \psi ^\dag \, [\hat{H},\, \sigma _\mu ]\, \psi\,,
\end{eqnarray}
where index $\mu =x,y,z$ refers to the spin polarization. It leads to the macroscopic equation of the spin density dynamics
\begin{eqnarray}
	\label{eq:dyn_spin_density}
	\frac{\partial S_\mu }{\partial t}+{\rm div}\, {\bf J}_\mu=-R_\mu\,,
\end{eqnarray}
where (index $i=x,y$ as we consider the 2D system)
\begin{eqnarray}
	\label{eq:Jmui}
	J_{\mu i}=\delta _{\mu i}\, \frac{v}{2\hbar }\, \psi ^\dag \psi
\end{eqnarray}
is the spin current and
\begin{eqnarray}
	\label{eq:Rmu}
	R_\mu=\frac{\ii v}{2\hbar }\, \epsilon _{\mu i\nu }\,
			\Big[ (\nabla _i\psi ^\dag )\sigma _\nu \psi -\psi ^\dag \sigma _\nu (\nabla _i\psi )\Big] \nonumber \\
		  -\frac{gM_i}{\hbar}\, \epsilon _{i\mu \nu }\, \psi ^\dag \sigma _\nu \psi
\end{eqnarray}
describes the spin relaxation rate. We note that even Eq.~(\ref{eq:Jmui}) explicitly does not depend on the coupling constant
$g$, the coupling between conductance electrons and localized spins is still preserved in the wave functions.
As we see in Eq.~(\ref{eq:Jmui}), the spin current ${\bf J}_\mu ({\bf r})$ does not depend on the perturbation $\hat{H}_{int}$ and is always transmitted along the wavevector ${\bf k}$, just like in the case of TI without any helicoid. This does not contradict to the suppression of the spin polarization in the state with a given ${\bf k}$ because the effect of the  perturbation is the creation of a superposition of
${\bf k}$ and ${\bf k\pm Q}$ states, which does not really affect the spin current direction.\cite{inversion}

\begin{figure}
	\centering
	\includegraphics[width=0.98\columnwidth]{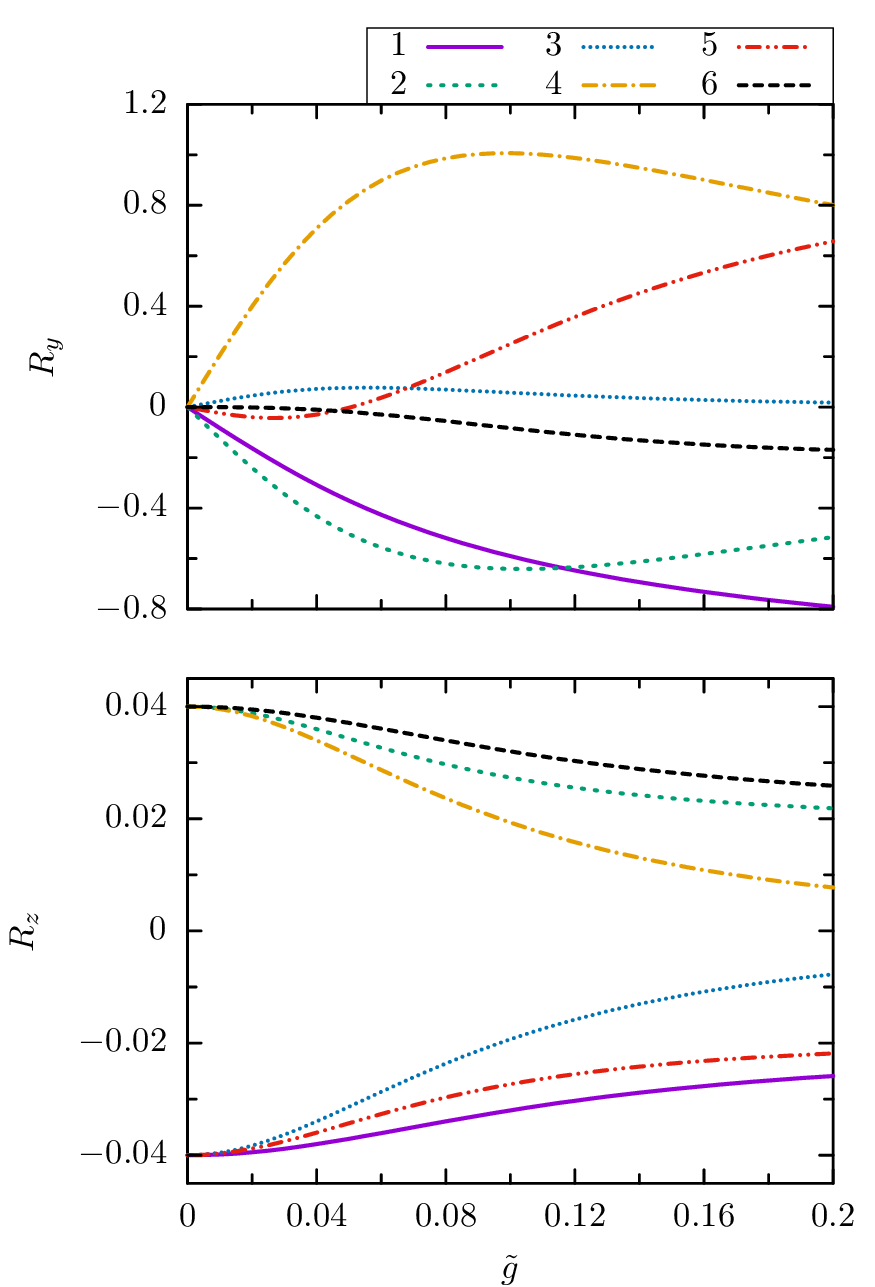}
	\caption{The relaxation terms $R_y$ and $R_z$ for fixed values of $\tilde{k}_x = 0.02$, $\tilde{k}_y = 0.0$, and $\tilde{Q}=0.2$.}
	\label{fig:RyRz}
\end{figure}

If the coupling $g=0$, the relaxation $R_\mu ({\bf r})$ is nonzero only for the  orientation of the spin perpendicular to the wave vector ${\bf k}$.
This result is understandable since for free TI electrons only the spin polarization parallel to the wave vector is allowed.
The perturbation related to the helicoid magnetization induces relaxation of all components of the spin density. The dependence of the spin relaxation rate as a function of coupling $g$ is presented in Fig.~\ref{fig:RyRz}. Different curves correspond to
different possible eigenstates of the perturbed Hamiltonian. Note that negative $R_{\mu}$ is acting on the system like a spin pumping term. For example, when the spin density is lower compared to the reference equilibrium value, then due to the negative$R_{\mu}<0$ the spin density is larger - relaxing to the equilibrium value.

\section{Spin torque-induced force}

The spin currents in a topological insulator are related to the free motion of electrons transmitting an angular momentum along the wave vector ${\bf k}$. Due to the coupling to the magnetic moments at the TI surface a spin torque is generated, which acts on the moments. Therewith one can expect creation of a resulting spin-current-induced mechanical force acting on the magnetic helicoid. As a result, one can observe the motion of the helicoid, which can be experimentally realized. Since magnetic moments are pinned, motion of the helicoid means a drift of the magnetic texture. Such a drift occurs because of a rotation of pinned magnetic moments.
The mechanical force can be calculated by using the standard relation ${\bf F}=\partial L/\partial {\bf r}$, where $L$ is the Lagrangian of the system.
In our case of TI with a magnetic helicoid, the Hamiltonian $\hat{H}$ depends on ${\bf r}$ in the direction of the helicoid vector ${\bf Q}$. Thus we have to calculate the quantity
\begin{eqnarray}
	\label{eq:Fx}
	F_x=-\psi ^\dag \; \frac{d\hat{H}_{int}}{dx}\; \psi\,,
\end{eqnarray}
which is the mean macroscopic $x$-component of the force. Then using Eq.~(\ref{eq:Hint}) for the perturbation
$\hat{H}_{int}$ we find
\begin{eqnarray}
	\label{eq:Fx_g}
	F_x=gM_0Q\, \psi ^\dag \big[ \sigma _x\sin (Qx)-\sigma _y\cos (Qx)\big] \psi\,.
\end{eqnarray}
Using Eq.~(\ref{eq:Fx_g}) and the eigenfunctions (\ref{eq:psi_nkr}) one can find the force induced by a single state $(n,{\bf k})$.
Taking the sum over all occupied electron states of the system one obtains the total force.

Clearly, the total force in the equilibrium of the system is zero. The reason is that in equilibrium we have equal number of electrons moving in opposite directions. Correspondingly, there is equal number of electrons with opposite spin orientations. Therefore there is no net force (force related to electron moving along $x$ is compensated by the force related to another one moving along  $-x$). For an applied external electric field a certain imbalance appears. This can be confirmed by direct calculations using the wave functions (\ref{eq:psi_nkr}). Therefore, we consider the linear response to a weak electric field $\boldsymbol{\mathcal{E}}$, taking the direction of electric field along axis $x$.

In the linear response approximation the total force can be presented by the following equation corresponding to the simple loop diagram (an analogue to the Kubo formula for the conductivity\cite{Nagaosa2010,Crepieux2001})
\begin{eqnarray}
	\label{eq:Fx_tot}
	F_x^{tot}=-\frac{ev\mathcal{E}}{2\pi } \, {\rm Tr}\,
	%\int \frac{d^2{\bf k}}{(2\pi )^2}\,
	\big( \hat{F}_x\, \hat{G}^R\, \sigma _x\, \hat{G}^A\big)\,,
\end{eqnarray}
where $\hat{F}_x=gM_0Q\, [\sigma _x\sin (Qx)-\sigma _y\cos (Qx)]$ is the force operator and
\begin{eqnarray}
	\label{eq:GR,A}
	\hat{G}^{R,A}=\big( \mu -\hat{H}\pm \ii\Gamma \big) ^{-1}
\end{eqnarray}
are, respectively, retarded and advanced Green's functions of the TI electronic states computed using the Hamiltonian $\hat{H}$; $\mu $ is the chemical potential and $\Gamma $ is the quantum decay rate of states related to the scattering from impurities.

\begin{figure}[ht]
	\includegraphics[width=0.98\columnwidth]{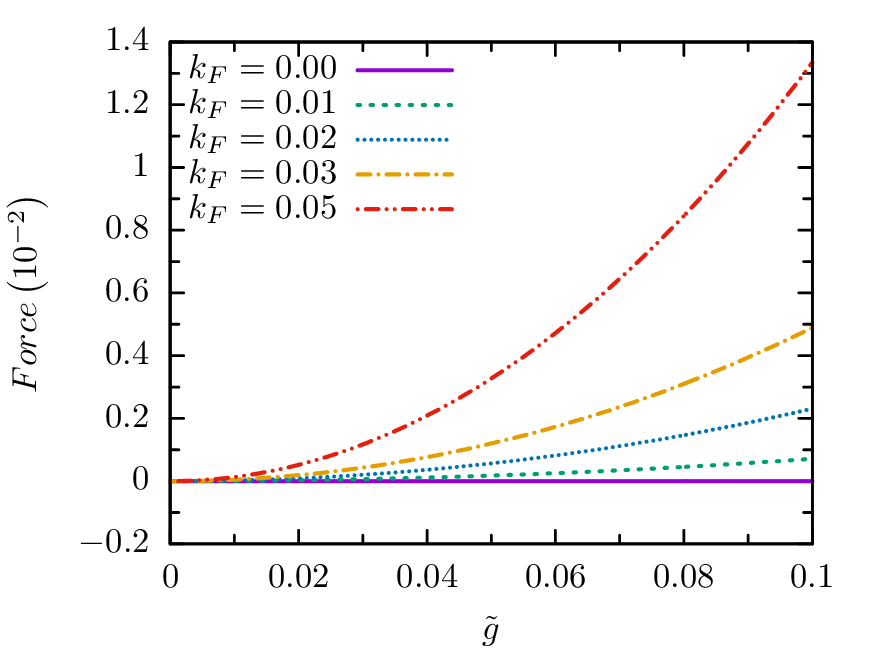}
	\caption{The dependence of total force on the coupling $\tilde{g}$ for $\tilde{Q} = 0.5$ and $\delta_k = 0.002$. The total force ${F}^{tot}_x = \tilde{F}^{tot}_x F_0$, where $F_0= g M_0 Q e \mathcal{E} / \left( 2 \pi v \right)$.}
	\label{fig:kF}	
\end{figure}

Using Eq.~(\ref{eq:Fx_tot}) we calculated the force $F_x^{tot}$ numerically. For this purpose all the operators in this equation have been
presented as matrices in the basis of the functions (\ref{eqs:phi_k}), whereby the trace in Eq.~(\ref{eq:GR,A}) includes 2D integration over the wave vector ${\bf k}$. The result of calculation of $F_x^{tot}$ as a function of the coupling $\tilde{g}$ is presented in Fig.~\ref{fig:kF} for different values of the chemical potential $\mu $. The magnitude of the force increases with the coupling and becomes larger with increasing the chemical potential $\mu $ since the transfer of the spin torque is due to the free carriers. Therefore, there is no force for $\mu =0$. Our calculations also show that the force is stronger for smaller value of $\tilde{Q}$. Clearly, the force is zero in the limit when the electron wave length is of the order of the helicoid period $L$, i.e., in the limit of $\tilde{Q}\sim k_F$.

\section{Conclusions}

The main purpose of the present work was to study the influence of the internal magnetic effects (magnetic impurities or the proximity of the single phase chiral MF system) on the electronic properties of topological insulators, which are known for the robustness of their electron energy spectrum. A symmetry breaking opens a gap in the energy spectrum. Hence, the band gap opening in the electronic spectrum by an applied external magnetic field is obvious, but "internal" magnetic effects may have subtle consequences.
While a single magnetic impurity is not able to alter the electronic spectrum, magnetic order in the system of localized spins has a number of important consequences. Here, we studied the case when localized magnetic impurities form a chiral spin order. We performed first principle density functional theory calculations and showed that the magnetic moments of Fe deposited on the surface of Bi$_2$Se$_3$ topological insulator form a chiral spin order. A very similar effect can be achieved by chiral multiferroics (a prototype material is  LiCu$_{2}$O$_2$) lodged in the proximity of TI. Our theoretical approach applies to both cases. We observed that, in contrast to the case of ferromagnetic ordering, in the case of the chiral spin order the Dirac point survives. An the energy gap appears at the edges of the new Brillouin zone. Another interesting result concerns the spin dynamics. We derived an equation for the spin density dynamics including a spin current and relaxation terms. We have shown that the motion of a conductance electron generates a magnetic torque and exerts a certain force on the helicoidal texture.

\begin{acknowledgments}
This work was supported by the National Science Center in Poland as a research project No.~DEC-2012/06/M/ST3/00042
and the DAAD under the PPP programme. We acknowledge financial support from DFG through priority program SPP1666 (Topological Insulators), SFB~762, and BE~2161/5-1. AK is supported by the Heisenberg Programme of the Deutsche Forschungsgemeinschaft (Germany) under Grant No.~KO~2235/5-1.
\end{acknowledgments}

\appendix

\section{The matrices of the spin operator}

Here we provide the matrices employed in the calculations
of Section VII B.
\label{app:S}
 \begin{gather}	
	 \label{mat:Szk}
	  S_{z{\bf k}}=  \left(
	     \begin{array}{cccccc}
	       0 & 0 & 0 & 1 & 0 & 0\\
	       0 & 0 & 0 & 0 & 1 & 0\\
	       0 & 0 & 0 & 0 & 0 & 1\\
	       1 & 0 & 0 & 0 & 0 & 0\\	
	       0 & 1 & 0 & 0 & 0 & 0\\
	       0 & 0 & 1 & 0 & 0 & 0
	     \end{array}
	    \right).
 \end{gather}
 \begin{gather}
		   S_{x{\bf k}}=\left(
		   \begin{array}{cccccc}
		    \frac{k_x}{|{\bf k}|} 	 & 0 				     & 0 				     & -\frac{\ii k_y}{|{\bf k}|} & 0	  & 0\\
		     0	         		 & \frac{k_x+Q}{\left|{\bf k+Q}\right|}   & 0 				     & 0 					 & -\frac{\ii k_y}{|{\bf k+Q}|} & 0\\
		     0	      			 & 0 			 	     & \frac{k_x-Q}{|{\bf k-Q}|}   & 0 					 & 0	  & -\frac{\ii k_y}{|{\bf k-Q}|}\\
		     \frac{\ii k_y}{|{\bf k}|} & 0				     & 0			   		 & -\frac{k_x}{|{\bf k}|} 	& 0	  & 0 \\
		     0	    			 & \frac{\ii k_y}{|{\bf k+Q}|} & 0 				     & 0 & -\frac{k_x+Q}{|{\bf k+Q}|}   & 0\\
		     0	    			 & 0 					 & \frac{\ii k_y}{|{\bf k-Q}|} & 0 & 0			 	      & -\frac{k_x-Q}{|{\bf k-Q}|}\\
		   \end{array}
		  \right).\nonumber\\
		  \label{mat:Sxk}
		\end{gather}
\begin{widetext}
		 \begin{gather}
		  S_{y{\bf k}}= \left(
		   \begin{array}{cccccc}
		    \frac{k_y}{|{\bf k}|} 	 & 0 				     & 0 				     & \frac{\ii k_x}{|{\bf k}|} & 0	  & 0\\
		     0	         		 & \frac{k_y}{|{\bf k+Q}|}   & 0 				     & 0 					 & \frac{\ii (k_x+Q)}{|{\bf k+Q}|} & 0\\
		     0	      			 & 0 			 	     & \frac{k_y}{|{\bf k-Q}|}   & 0 					 & 0	  & \frac{\ii (k_x-Q)}{|{\bf k-Q}|}\\
		     -\frac{\ii k_x}{|{\bf k}|} & 0				     & 0			   		 & -\frac{k_y}{|{\bf k}|} 	& 0	  & 0 \\
		     0	    			 & -\frac{\ii (k_x+Q)}{|{\bf k+Q}|} & 0 				     & 0 & -\frac{k_y}{|{\bf k+Q}|}   & 0\\
		     0	    			 & 0 					 & -\frac{\ii (k_x-Q)}{|{\bf k-Q}|} & 0 & 0			 	      & -\frac{k_y}{|{\bf k-Q}|}\\
		   \end{array}
		  \right).\nonumber\\
		  \label{mat:Syk}
		\end{gather}
\end{widetext}

\end{document}